\documentclass[a4paper, usenatbib]{mnras}

\usepackage[T1]{fontenc}
\usepackage{ae,aecompl}

\usepackage{graphicx}	
\usepackage{amsmath}	
\usepackage{amssymb}	
\usepackage{lineno}
\usepackage{gensymb}
\usepackage{caption}
\usepackage{textcomp}
\usepackage{lipsum}
\usepackage{hyperref}
\usepackage{natbib}
\usepackage{url}
\usepackage{float}
\usepackage{upgreek}
\usepackage{lscape}
\usepackage{multirow}
\usepackage{array}
\usepackage{soul}
\usepackage[utf8]{inputenc}

\title[Dwarf spheroidal J-factors without priors]{Dwarf spheroidal J-factors without 
priors: A likelihood-based analysis for indirect dark matter searches}

\author[Chiappo et al.]{
A. Chiappo$^{1,2}$,\thanks{E-mail: andrea.chiappo@fysik.su.se}
J.Cohen-Tanugi$^{3}$,
J. Conrad$^{1,2}$,\thanks{Wallenberg Academy Fellow}
L. E. Strigari$^{4}$,
B. Anderson$^{1,2}$,
\newauthor M.A. S\'anchez-Conde$^{1,2}$
\\
$^{1}$The Oskar Klein Centre for Cosmoparticle Physics, AlbaNova, SE-106 91 Stockholm, Sweden\\
$^{2}$Department of Physics, Stockholm University, AlbaNova, SE-106 91 Stockholm, Sweden\\
$^{3}$Laboratoire Univers et Particules de Montpellier, IN2P3/CNRS et  Universit\'e de 
Montpellier, 34095 Cedex 05 Montpellier, France\\
$^{5}$Mitchell Institute for Fundamental Physics and Astronomy,  Department of Physics 
and Astronomy, Texas A\&M University,\\ College Station, TX 77845, USA
}

\date{Accepted 2016 November 24. Received 2016 November 11; in original form 2016 August 
25}

\pubyear{2016}

\begin{document}
\label{firstpage}
\pagerange{\pageref{firstpage}--\pageref{lastpage}}
\maketitle

\begin{abstract}
Line-of-sight integrals of the squared density, commonly called the \textit{J-factor}, are 
essential for inferring dark matter (DM) annihilation signals. The J-factors of 
DM-dominated dwarf spheroidal satellite galaxies (dSphs) have typically been derived using 
Bayesian techniques, which for small data samples implies that a choice of priors 
constitutes a non-negligible systematic uncertainty. Here we report the development of a 
new fully frequentist approach to construct the profile likelihood of the J-factor. Using 
stellar kinematic data from several classical and ultra-faint dSphs, we derive the 
maximum likelihood value for the J-factor and its confidence intervals. We validate this 
method, in particular its bias and coverage, using simulated data from the \textit{Gaia 
Challenge}. We find that the method possesses good statistical properties. The J-factors 
and their uncertainties are generally in good agreement with the Bayesian-derived values, 
with the largest deviations restricted to the systems with the smallest kinematic data 
sets. We discuss improvements, extensions, and future applications of this technique.
\end{abstract}



\section{Introduction}

There exists now a wide range of particle physics models extending beyond the Standard 
Model of particle physics which accommodate dark matter (DM) candidates. One of the most 
studied ones features a weakly interactive massive particle (WIMP) (for recent reviews 
see \citealt{Bertone:2004pz}, \citealt{Feng:2010gw}, \citealt{Conrad:2014tla}), which is 
expected to yield gamma rays through decay or pair annihilation. The differential photon 
flux expected from DM annihilation is given by
\begin{equation}
  \frac{\text{d}N_\gamma}{\text{d}E_\gamma} = \frac{1}{4\uppi}  
\underbrace{\frac{\langle\sigma v\rangle}{2m^2_\text{DM}} \sum_i 
B_i\frac{\text{d}N_i(E)}{\text{d}E}}_\text{particle physics factor} \, \times \, \text{J} 
\quad .
 \label{eq1}
\end{equation}
Equation (\ref{eq1}) describes the dependence of the photon flux on the properties of the 
candidate particle, such as its mass, $m_\text{DM}$, its velocity-averaged thermal 
cross-section, $\langle\sigma v\rangle$, and the sum of photons in each possible final 
state $i$, which has branching ratio $B_i$ and photon yield $\text{d}N_i(E)/\text{d}E$. 
The last term, the \textit{J-factor}, is the line-of-sight (l.o.s.) integral of 
the DM halo density $\rho_\text{DM}$ squared (in the case of annihilation), integrated 
over a solid angle $\Delta\Omega = 2\uppi (1-\text{cos}\,\theta_\text{max})$. It is given 
by
\begin{equation}
\text{J}(D,\Delta\Omega)= \int_{\Delta\Omega} \int_\text{l.o.s.} \rho^2_\text{DM}(r(s)) 
\text{d}s \text{d}\Delta\Omega' .
\label{eq2}
\end{equation}
Because of this square density dependence of the flux, regions of density enhancements 
are targets for DM searches. For instance, J $\approx$ $10^{22} \mbox{-} 10^{23}$ GeV$^2$ 
cm$^{-5}$ for the Galactic centre (GC), $10^{17} \mbox{-} 10^{19}$ GeV$^2$ cm$^{-5}$ for 
dwarf galaxies, and $10^{15} \mbox{-} 10^{19}$ GeV$^2$ cm$^{-5}$ for galaxy clusters (see 
\citealt{Conrad:2015bsa} and \citealt{Charles:2016pgz} for a discussion of different 
targets for DM searches). These values would place the GC at the top of the candidate 
targets, however the presence of strong, yet unmodelled fore- and background 
contaminations renders it a very challenging target (see, for example, 
\citealt{Zhou:2014lva}, \citealt{Calore:2014xka}, \citealt{TheFermi-LAT:2015kwa}). For 
this reason and due to the nearly negligible presence of such contaminations, dSphs are 
considered the ideal targets for gamma-ray DM searches.

Assuming that DM uniquely consists of WIMPs, the photons resulting from their 
annihilations within DM haloes should have energies in the gamma-ray range, hence 
detectable by instruments such as the \textit{Fermi} Large Area Telescope 
\citep[LAT]{2009ApJ...697.1071A} or the ground based IACTs like HESS, VERITAS or MAGIC 
(\citealt{Zitzer:2015uta}, \citealt{Ahnen:2016qkx}, \citealt{Giammaria:2016jfo}, 
\citealt{Mora:2015vhq}). Indeed, in recent years many groups have used the LAT data in 
direction of dSphs to search for $\gamma$-ray signals (\citealt{Ackermann:2011wa}, 
\citealt{Ackermann:2015zua}, \citealt{Geringer-Sameth:2014qqa}, 
\citealt{Drlica-Wagner:2015xua}).

In these analyses equation (\ref{eq1}) is used to extract limits on $\left<\sigma v 
\right>$ from the data for a range of $m_\text{DM}$. The viable DM masses are usually 
suggested by standard model extensions (SMEs), which also give the spectrum of the 
possible annihilation channels $i$ (for a recent review on the various SMEs see 
\citealt{daSilva:2014qba}). The remaining quantity to be determined is the J-factor; this 
term represents the major systematic uncertainty in DM searches. The now-standard 
technique for determining the J-factor involves a likelihood-based fitting to the stellar 
photometric and kinematic data within a Bayesian framework. The stellar and DM 
distributions are typically assumed to be spherically-symmetric, but possibly with 
anisotropic stellar velocity distribution. A Markov Chain Monte Carlo (MCMC) is used to 
sample the posterior probability distribution function of the J-factor for a given choice 
of the parameters assumed in $\rho_\text{DM}$~(\citealt{Martinez:2009jh}, 
\citealt{2015MNRAS.451.2524M}, \citealt{2015MNRAS.446.3002B}, 
\citealt{Geringer-Sameth:2014yza}, \citealt{Bonnivard:2015pia}, \citealt{Ullio:2016kvy}). 
However, in cases of small stellar data samples, the determination of the J-factor is 
subject to the choice of theoretical priors \citep{Martinez:2009jh}. This shortcoming is 
not present in a frequentist treatment of the problem.

From the perspective of kinematic modelling of the data, several authors have recently 
proposed improvements upon this standard spherically-symmetric analysis, and have explored 
the implications for current and future gamma-ray observations.~\cite{2015MNRAS.446.3002B} 
use mock kinematic data with underlying triaxial DM haloes, and analyse the data under the 
assumption of the spherically-symmetric Jeans equation. They identify biased J-factor 
estimates for the mildly triaxial halo models they consider. ~\citet{Hayashi:2016kcy} 
consider an axisymmetric Jeans model.~\citet{Sanders:2016eie} consider a flattened model 
for the stellar and DM distribution, and estimate the correction for the J-factor relative 
to spherical models for prolate and oblate DM haloes. Each of these models that utilize 
the axisymmetric Jeans equations requires model-dependent assumptions on the 
three-dimensional shape of the DM density distribution, and where necessary assume a 
Gaussian likelihood function for the stellar velocities.

In this paper we adhere with the assumption of spherical symmetry, but differ from the 
standard modelling by presenting the first fully frequentist derivation of the dSphs 
J-factor based on the profile likelihood technique. This approach has two main 
advantages: first, it avoids the additional systematic uncertainty introduced by the 
choice of priors in a Bayesian analysis and secondly, it allows a more consistent 
treatment of J-factor uncertainties in gamma-ray analyses which mostly are frequentist. We 
explicitly compare and contrast to the results from previous Bayesian analyses.

This paper is organized as follows. Section~\ref{sec2} describes our analysis method in 
detail. Section~\ref{sec3} presents a validation of this technique based on the 
simulations produced by the \textit{Gaia} Challenge team. In Section~\ref{sec4} we 
calculate the J-factor for 20 dSphs, compare our results to previously obtained values 
available in the literature, and also discuss the properties of our results and the 
validation. We conclude with a discussion on future prospects for DM searches and possible 
improvements of the technique.

\section{Method}
\label{sec2}
For the l.o.s. stellar velocity data we assume a Gaussian likelihood function of the form 
(\citealt{Walker:2005nt}, \citealt{Strigari:2008ib}):
\begin{equation}
\mathcal{L} = -\text{log}L = \frac{1}{2}\sum^{N_\star}_{i=1} 
\left[\frac{(v_i-u)^2}{\sigma^2_i}+\text{log}(2\uppi\sigma^2_i)\right] \quad ,
 \label{eq3}
\end{equation}
\noindent where the index $i$ runs over each of the $N_\star$ stars in the sample, and 
where $v_i$ and $u$ are the particular and mean velocity, respectively. The expected 
velocity dispersion $\sigma_i$ is taken as the squared sum of a measurement uncertainty 
$\epsilon_i$ and a l.o.s. systemic dispersion $\sigma_\text{los}(R_i)$, which depends on 
the projected radial distance of star $i$ to the centre of the system:  $\sigma^2_i = 
\epsilon^2_i+\sigma^2_\text{los}(R_i)$. We note that the Gaussian likelihood function in 
equation (\ref{eq3}) is likely an approximation to the probability distribution for a 
true stellar velocity at a projected position, which is ultimately a function of the 
dynamical model for the stars and the DM. For dSphs in which the true velocity dispersion 
is similar to the dispersion from the measurement uncertainty, such as ultra-faint dwarfs, 
$\sigma_i$ is dominated by Gaussian measurement uncertainty, so equation (\ref{eq3}) is a 
good approximation. 

The second assumption of our analysis, which is again standard, is the spherical Jeans 
equation~\citep{2008gady.book.....B}. This is used to link  $\sigma^2_\text{los}(R_i)$ to 
the underlying gravitational potential, assumed to be entirely defined by the DM halo 
density profile. The solution of the spherical Jeans equation for the l.o.s. velocity 
dispersion is 
\begin{equation}
\begin{aligned}
\sigma^2_\text{los} (R) & = \frac{2}{I(R)} \int^\infty_R 
\left(1-\beta(r)\frac{R^2}{r^2}\right) 
 \frac{r}{\sqrt{r^2-R^2}} \frac{1}{f(r)} \\
 & \times \int^\infty_r f(s) \frac{\nu_\star(s)\text{GM}(s)}{s^2} 
\text{d}s\,\text{d}r\quad ,
\label{eq4}
\end{aligned}
\end{equation}
\noindent with
\begin{equation}
 f(r) = f_{r^\prime} \text{exp}\left[\int^{r}_{r^\prime} 
\frac{2\,\beta(t)}{t}\text{d}t\right]\quad .
\end{equation}
Here, $R$ is the projected distance from the star to the centre of the stellar system; 
$M(s) = 4\uppi\int^s_0 \rho_{DM}(r)r^2\text{d}r$ is the total enclosed DM mass; 
$\nu_\star$ is the density profile of the luminous component of the system, $I(R)$ its 
corresponding surface brightness, and $\beta(r)$ is the velocity anisotropy profile.

Inferring $\rho_\text{DM}(r)$ from stellar data, usually in the form of a parametrized 
function, directly allows for the computation of the J-factor in equation (\ref{eq2}). In 
the literature, this inference proceeds via a Bayesian analysis which samples (via MCMC 
for instance) a posterior probability density built from equation (\ref{eq3}) and a prior 
density that captures further information (\citealt{Martinez:2009jh}, 
\citealt{Geringer-Sameth:2014yza}, \citealt{Bonnivard:2015pia}). Marginalization of the 
parameters eventually yields the J-factor probability density by direct integration. 

Alternatively, it is straightforward to derive a likelihood function dependent on the 
J-factor directly, and to draw inference from it. To do so, we note that the 
parametrized functions defining  $\rho_\text{DM}(r)$ are generically first order 
polynomials in a scale density parameter $\rho_0$, while also depending on a scale radius 
$r_0$, so that one can write $M(s, r_0, \rho_0)=\rho_0\,m(s, r_0)$ and $J(r_0, \rho_0, D, 
\Delta\Omega)=\rho_0^2\,j(r_0, D, \Delta\Omega)$. Then, simply replacing $M(s, r_0, 
\rho_0)$ in equation (\ref{eq4}) by $\sqrt{J/j(r_0, D, \Delta\Omega)}\,m(s, r_0)$ yields 
the 
desired formalism for a direct likelihood analysis on the parameter $J$.

To completely determine the problem, one needs to adopt a specific functional expression 
for $\rho_\text{DM}(r)$, $\nu_\star(r)$ and $\beta(r)$. Here we use the generalized 
Hernquist profile~\citep{Hernquist:1990be}:
\begin{equation}
 \nu_\star (r) = 
\rho_\star\left(\frac{r}{r_\star}\right)^{-\gamma}\left(1+ 
\left(\frac{r}{r_\star}\right)^{\alpha}\right)^{-\frac{\beta-\gamma}{\alpha}}\quad
,
 \label{eq6}
\end{equation}
for both $\rho_\text{DM}$ and $\nu_\star$ (with the `$\star$' subscript replaced by `$0$' 
for $\rho_\text{DM}$), with the following prescription. The DM is either a cusped 
or cored Zhao profile \citep{Zhao:1995cp}, corresponding to $(\alpha, \beta,\gamma)=(1,3, 
1)$ or $(1,3,0)$; the former corresponds to the well-known NFW profile 
\citep{Navarro:1996gj}. For the light profile $I(R)$, we make use of either a Plummer 
\citep{1911MNRAS..71..460P}, Plummer-like, or non-Plummer profiles, which are obtained 
via the Abel transform\footnote{The expression of the Abel transform for the projection 
of 
a quantity $f(r)$ into $F(R)$ reads $F(R) = 2 \int^{+\infty}_R \frac{f(r) r 
\text{d}r}{\sqrt{r^2 - R^2}}$, where $R$ is the projected radius.} of equation 
(\ref{eq6}) 
with $(\alpha, \beta,\gamma)=(2, 5, 0)$, $(2,5,0.1)$, or $(2,5,1)$, respectively. 
Observing equation (\ref{eq4}), we notice that $\nu_\star / I(R)$ is independent of the 
normalization parameter $\rho_\star$, which can thus be arbitrarily set to unity.

Finally, we consider here three parameterizations for $\beta(r)$: an isotropic stellar 
velocity distribution with $\beta(r) = 0$, a constant anisotropy with $\beta(r) = \beta$, 
and an Osipkov-Merritt profile (OM, \citealt{1979SvAL....5...42O}, 
\citealt{1985AJ.....90.1027M})
\begin{equation}
 \beta(r) = \frac{r^2}{r^2+r_a^2} \quad. 
 \label{eq7}
\end{equation}
In this formula $r_a$ is the scale radius at which the velocity profile shifts from 
isotropic to anisotropic. The choice for the OM profile is further discussed below. We 
are interested in the maximum likelihood estimate (MLE) and likelihood curve for the 
J-factor, $J$, with the scale radius parameters, $r_0$, $r_\star$, and the value of 
$\beta$ or $r_a$ in the two last anisotropy scenarios playing the role of nuisance 
parameters. 

\begin{figure}
 \includegraphics[scale=0.45]{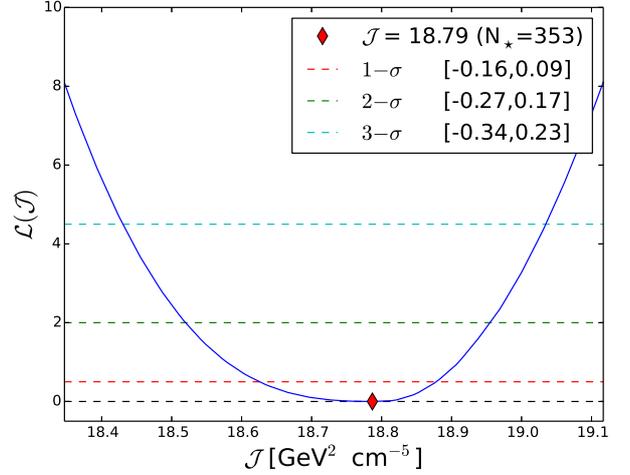}
 \caption{Profile likelihood of $\mathcal{J}$ obtained from the application of the MLE 
scheme on the stellar data of Draco (see text). An isotropic model for the stellar 
velocity distribution was assumed.}
 \label{fig1} 
\end{figure}

With these ingredients in mind, we fit for $J$ with a profile likelihood technique that 
also yields the likelihood curve. For each fixed value of $J$ scanned over a reasonable 
range, we minimize the resulting likelihood function 
$\cal{L}_\textit{J}(\vec{\text{R}},\vec{\text{v}},\vec{\epsilon}\,|\vec{\textit{g}})$ with 
respect to the nuisance parameters $\vec{\textit{g}}$, thus obtaining, by interpolation, 
the likelihood curve over $J$ together with its minimum. For data (Sec.~\ref{sec4}), 
$r_\star$ is always fixed to its value in the literature \citep{2012AJ....144....4M}, 
whereas for simulations (Sec.~\ref{sec3}), to its true value. In the notation above, the 
nuisance parameter vector $\vec{\textit{g}}$ always includes $r_0$, plus $\beta$ or $r_a$ 
depending on the anisotropic stellar velocity profile assumed. Fitting a different profile 
than simulated, or freeing structural parameters (e.g. the three exponents of the Zhao 
profile) is left for a future work. The vectors $(\vec{R}$, $\vec{v}$, $\vec{\epsilon})$ 
stand for the data arrays associated with each dSph (radius, velocity, and velocity 
measurement uncertainty, respectively; see \citealt{2012AJ....144....4M}, 
\citealt{2013pss5.book.1039W}, \citealt{Battaglia:2013wqa}, \citealt{Strigari:2013iaa} for 
further reference on the dSphs stellar data). The minimization is performed with the 
MINUIT\footnote{We used the iminuit python implementation: 
\href{https://pypi.python.org/pypi/iminuit}{https://pypi.python.org/pypi/iminuit}.} 
library. Given the scale range of the J-factor that we need to consider, we actually fit 
the parameter $\mathcal{J}=\text{log}_{10}\text{(J)}$ instead of $J$. Fig.~\ref{fig1} 
shows the likelihood curve $\mathcal{L}(\mathcal{J})$ obtained with our procedure in the 
case of the Draco dSph, when assuming an isotropic stellar velocity distribution.

\begin{table*}
\caption{Models tested with the MLE scheme. For each, our code was run on the 100 and 
1000 star data sets. All models assume $r_0$ = 1 kpc.}
\label{tab1}
 \begin{tabular}{ l r c c c c }
   \hline
 Model & $\rho_0$ ($\text{M}_\odot \text{kpc}^{-3}$) & $\mathcal{J}$ (GeV$^2$ cm$^{-5}$) 
& $r_a$ (kpc) & $\gamma$ & $r_\star$ (kpc)\\
 \hline
 OM Cored non-Plummer 		&   4$\times 10^8$ & 19.23	& 0.25 		& 1 	
	& 0.25 	\\
 OM Cored Plummer-like		&   4$\times 10^8$ & 19.23 	& 0.25 		& 0.1 	
	& 0.25 	\\
 Isotropic Cored non-Plummer 	&   4$\times 10^8$ & 19.23 	& $\infty$	& 1 	
	& 1 	\\
 Isotropic Cored Plummer-like	&   4$\times 10^8$ & 19.23 	& $\infty$	& 0.1 	
	& 1	\\
 OM Cusped non-Plummer 		& 6.4$\times 10^7$ & 18.83	& 0.1		& 1 	
	& 0.1 	\\
 OM Cusped Plummer-like		& 6.4$\times 10^7$ & 18.83	& 0.1 		& 0.1 	
	& 0.1 	\\
 Isotropic Cusped non-Plummer 	& 6.4$\times 10^7$ & 18.83	& $\infty$ 	& 1 	
	& 0.25	\\
 Isotropic Cusped Plummer-like	& 6.4$\times 10^7$ & 18.83	& $\infty$ 	& 0.1 	
	& 0.25 	\\
 \hline
\end{tabular}
\end{table*}
 
\begin{figure*}
\centering
 \includegraphics[scale=0.35]{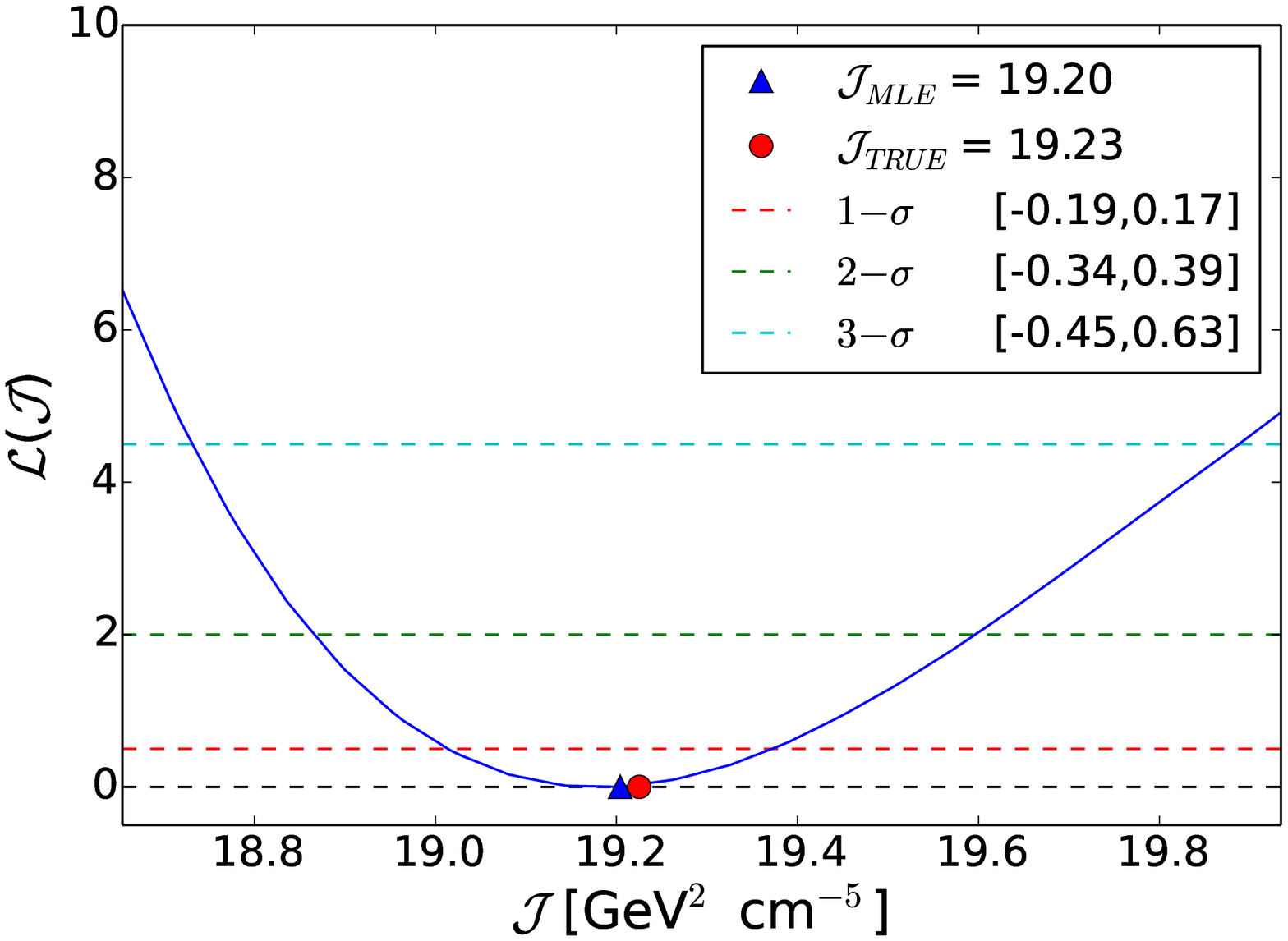}
 \includegraphics[scale=0.35]{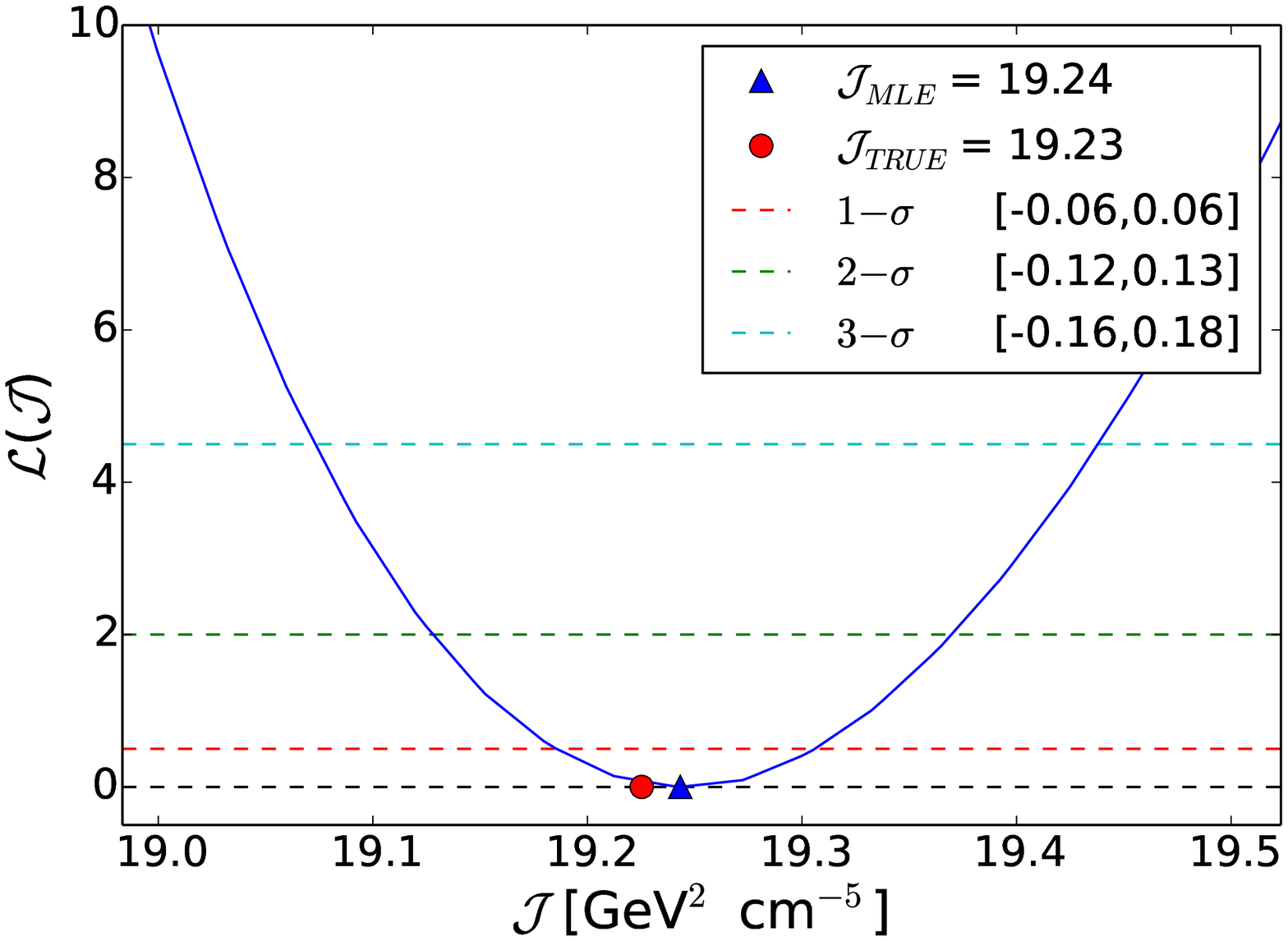}
 \caption{Results of the MLE scheme applied on the \textit{Gaia} Challenge mock data sets 
  of the Isotropic Cored Plummer-like model. The plot on the left (right) was obtained 
  using the 100 (1000) stars sample. The likelihood curves were obtained assuming the 
true 
  model in equation \ref{eq4} and profiling over the nuisance parameter array 
  $\vec{\textit{g}}$ (see text).}
 \label{fig2}
 \end{figure*}
 
\begin{figure*}
\centering
 \includegraphics[scale=0.35]{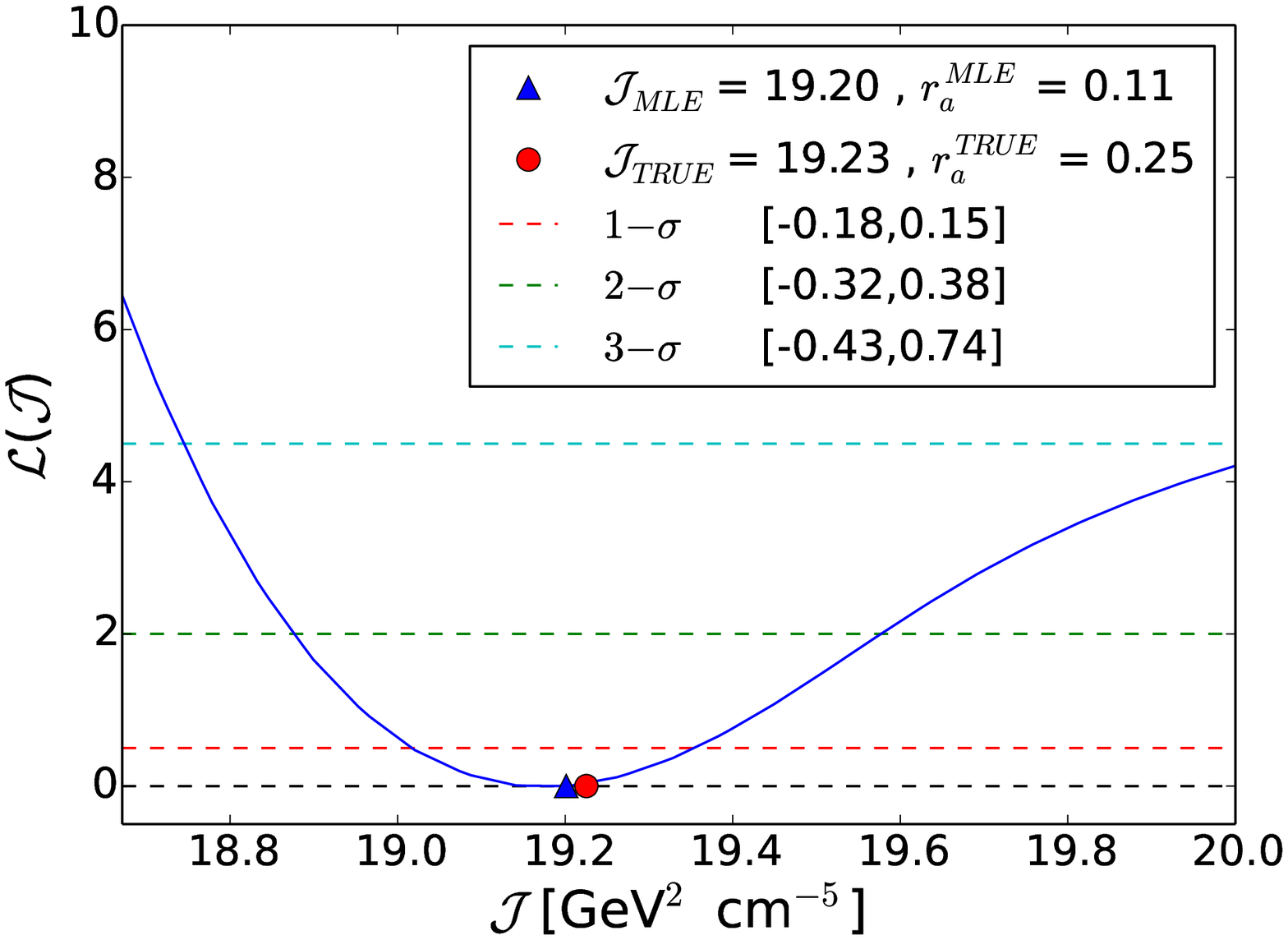}
 \includegraphics[scale=0.35]{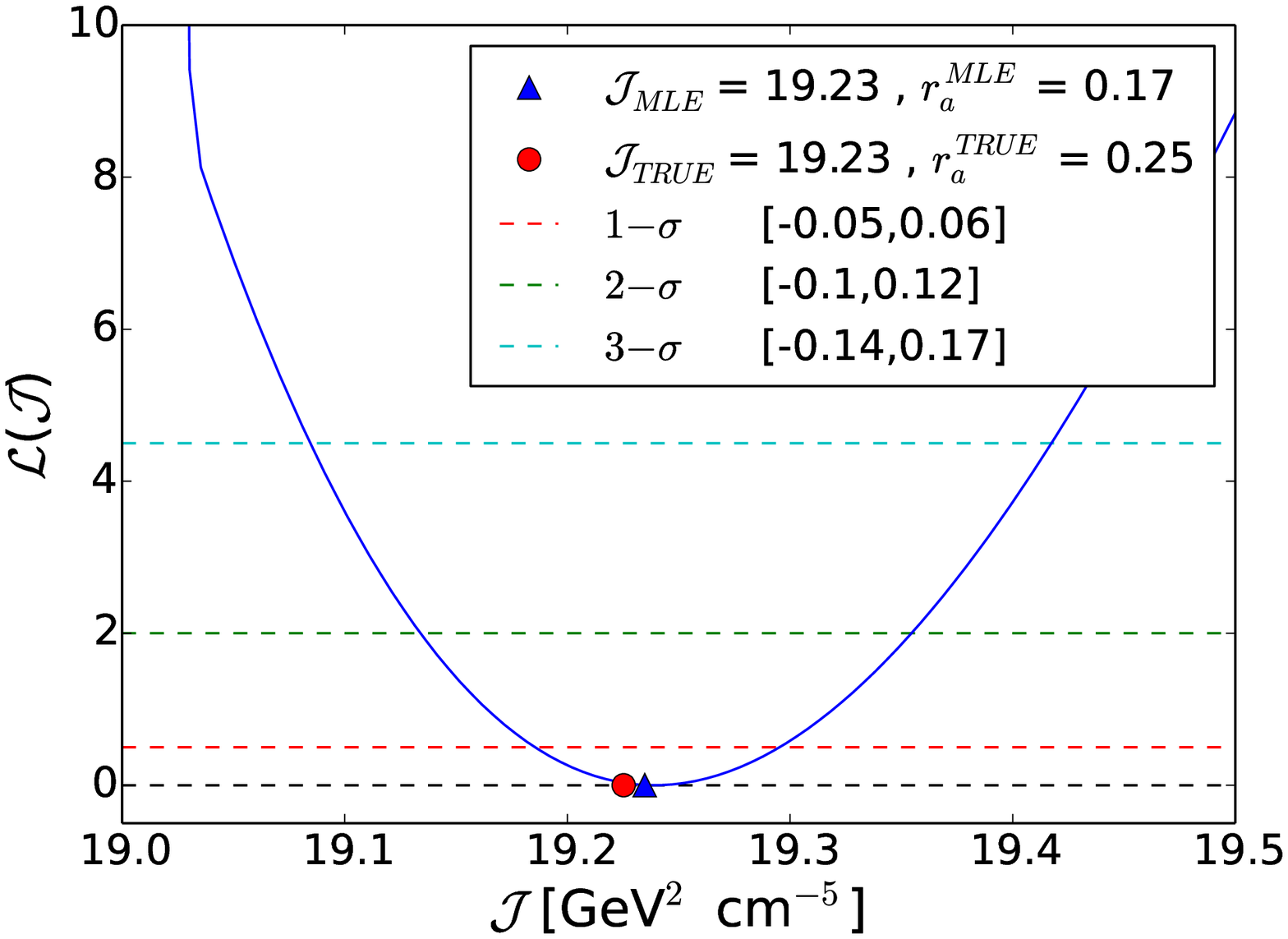}
 \caption{Results of the MLE scheme applied on the \textit{Gaia} Challenge mock data sets 
  of the OM Cored non-Plummer anisotropic model. The plot on the left (right) was 
obtained 
  using the 100 (1000) stars sample. The likelihood curves were obtained assuming the 
true 
  model in equation \ref{eq4} and profiling over the nuisance parameter array 
  $\vec{\textit{g}}$ (see text).}
 \label{fig3}
 \end{figure*}

\section{Validation}
\label{sec3}
In order to obtain $\cal{L}$($\mathcal{J}$) by applying the MLE scheme outlined in the 
previous section, we have developed a \textsc{python} \, code called 
\textsc{astrojpy}~\footnote{\textsc{astrojpy} \, will be soon released to the public. The 
interested reader is welcome to contact the authors.}. We validate the procedure by 
running our code on the simulated data produced by the \textit{Gaia} Challenge 
team~\footnote{\url{http://astrowiki.ph.surrey.ac.uk/dokuwiki/doku.php?id=workshop}} 
\citep{2011ApJ...742...20W}. These simulations are designed to mimic spherical and 
triaxial collision-less stellar systems belonging to, e.g., the Milky Way, its dSphs, or 
giant elliptical galaxies. They were generated to provide a verification tool for the 
mass modelling of the systems that the \textit{Gaia} satellite is currently observing 
\citep{2012Ap&SS.341...31D}. They consist of random samplings from a stellar distribution 
function model. Here we focus on spherically symmetric models, for the DM adopting cusped 
and cored Zhao profiles, and for the stars adopting the Plummer-like or non-Plummer 
profiles as is described above (Sec.~\ref{sec2}). \\
\indent The \textit{Gaia} Challenge simulations were generated from a stellar 
distribution function assuming an OM anisotropy profile, so for our mock data we work 
within the context of this model. In order to generate an isotropic model from these 
simulations, we consider the fiducial model with $r_a = \infty$.  Models with variable 
anisotropy are obtained by setting $r_a$ to a value similar to the characteristic scale 
in 
the light distribution, $r_\star$. \\
\indent As we would like the \textit{Gaia} Challenge simulations to resemble the 
properties of observed dSphs as much as possible, mock systems would be ideally built 
with 
velocity dispersions of $\approx 5-10\, \text{km s}^{-1}$, roughly constant at all 
projected radii from the centre of the dSphs. For example, models with substantial radial 
velocity anisotropies have observed velocity dispersions that vary much more than the 
corresponding profiles of the observed dSphs. Thus, in our choice of fiducial models, we 
must balance the fact that we are considering a specific OM anisotropy profile with the 
fact that we are constrained to mimic the actual dSph data. \\
\begin{figure*}
\centering
\includegraphics[scale=0.35]{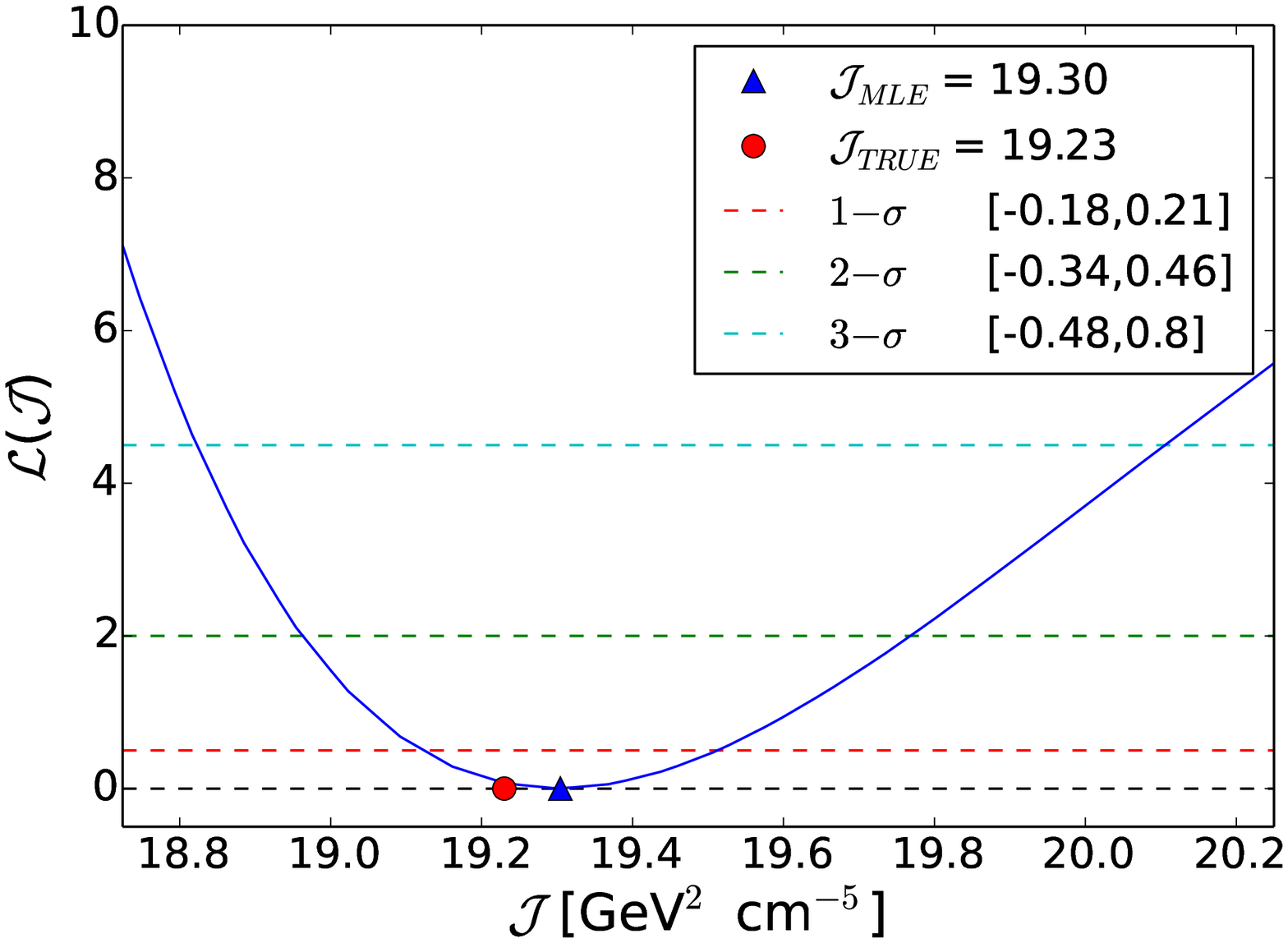}
\includegraphics[scale=0.35]{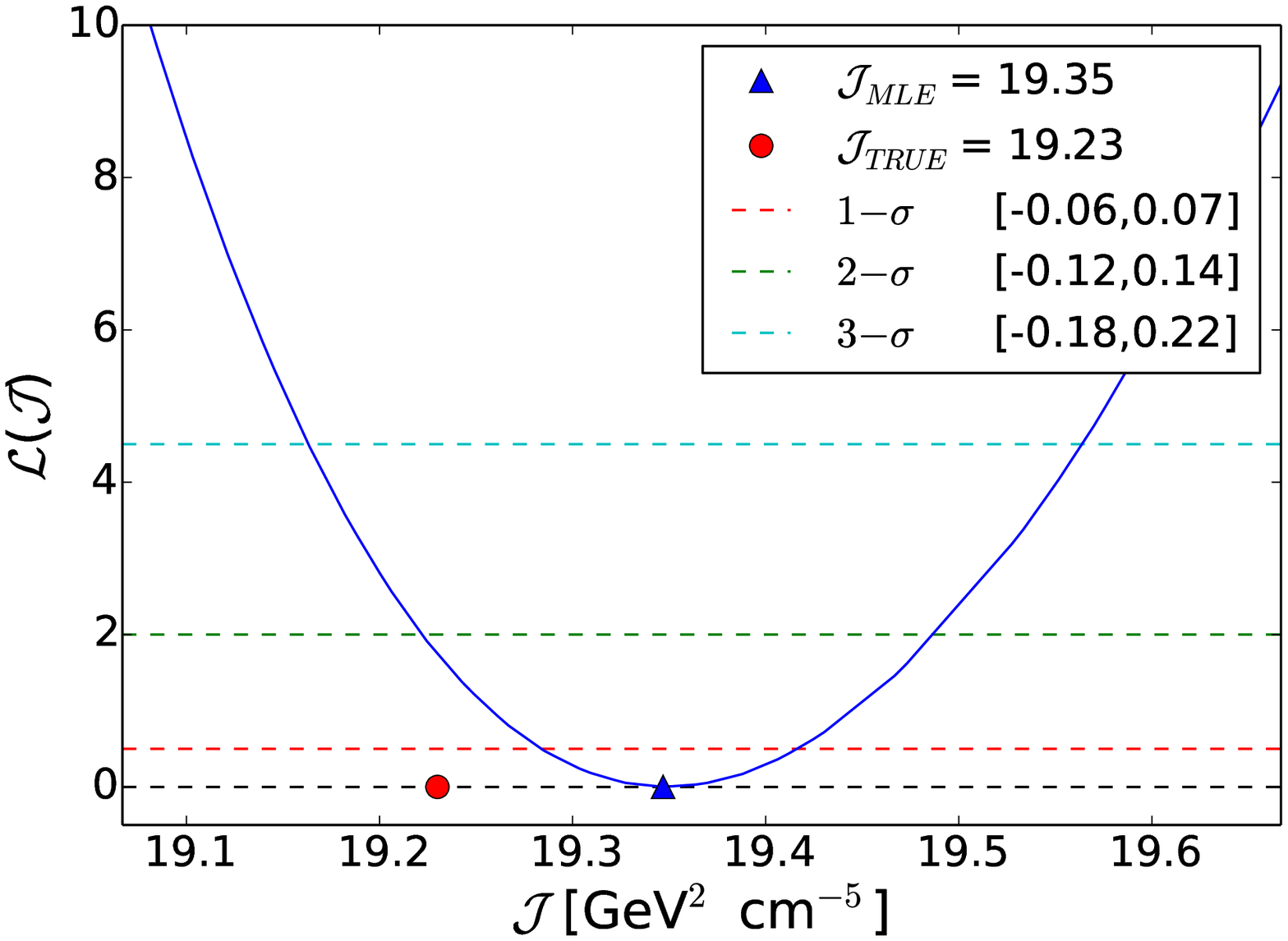}
 \caption{Results of the MLE scheme applied on the \textit{Gaia} Challenge mock data sets 
  of the Isotropic Cored Plummer-like model. The plot on the left (right) was obtained 
  using the 100 (1000) stars sample. The likelihood curves were obtained assuming the 
  Isotropic NFW Plummer model, thus the same model used on dSphs data, and profiling over 
  the nuisance parameter array $\vec{\textit{g}}$ (see text).}
 \label{fig4}
\end{figure*}
\begin{figure*}
\includegraphics[scale=0.35]{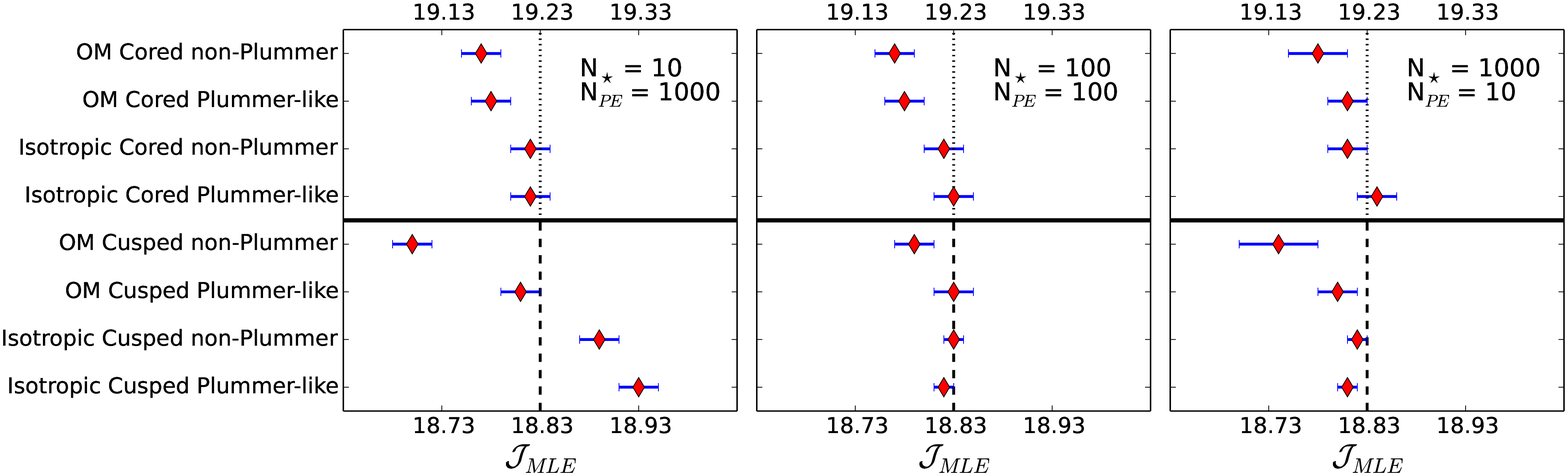}
 \caption{Bias estimates for the eight models provided by \textit{Gaia} Challenge. The 
  red dots represent the mean values of the $\mathcal{J}_\text{MLE}$ obtained in the 
  pseudo-experiments (PE) generated by partitioning the full data set (see text for 
  details). The errorbars are the 1$\sigma$ uncertainties on these means. The dashed 
  (dotted) vertical lines indicate $\mathcal{J}_\text{TRUE}$ of the cusped (cored) models 
  (see Table~\ref{tab1}).}
 \label{fig5} 
\end{figure*}
\indent Motivated by these considerations we choose a set of eight different models, 
which are shown in Table~\ref{tab1}. In our tests we use simulated data sets of 100 and 
1000 stars generated from the appropriate distribution functions. We project the 
three-dimensional sets of positions and velocities generated from the distribution 
function along their $z$-axis, and set the distance to the centre of the simulated dSphs 
to a value of $D = 100$ kpc. \\
\indent Assuming the true model in equation ({\ref{eq4}}), we let the code retrieve the 
true $\mathcal{J}$ factor ($\mathcal{J}_\text{TRUE}$), integrating up to 
$\theta_\text{max} = 0.5^\circ$ in equation (\ref{eq2}) and assuming the same distance 
$D$, while profiling over the nuisance parameters $\vec{\textit{g}}$. The results of the 
tests for one isotropic model and one anisotropic model are shown in Figs.~\ref{fig2} and 
\ref{fig3}. In both figures, the plots on the left (right) column were obtained using the 
100 (1000) stars data sets. We notice that the uncertainty in the best fitting values 
($\mathcal{J}_\text{MLE}$) consistently decreases with growing sample size, as evident 
from the narrowing of the confidence intervals. Moreover, $\mathcal{J}_\text{MLE}$ 
approaches $\mathcal{J}_\text{TRUE}$ as $N_\star$ grows, as seen by comparing the left 
plots with the right ones in Figs.~\ref{fig2} and \ref{fig3}. A similar trend occurs for 
the nuisance parameters, as shown, e.g., for the anisotropy parameter $r_a$ in the 
legends 
of Fig.~\ref{fig3}. The results for the other six models examined (not shown) exhibit a 
similar behaviour.
\begin{figure*}
 \includegraphics[scale=0.35]{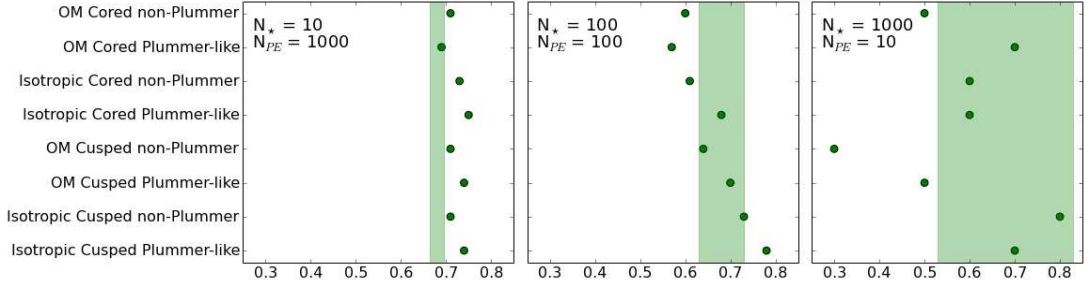}
 \caption{Results for the 1$\sigma$ coverage test. The points represent the coverage of 
  each of the eight models used in the validation ($y$-axis, see Table~\ref{tab1}). The 
  green bands represent the expected coverage of an ideal test, i.e. one which yields 
  exactly 68 per cent.}
\label{fig6}
\end{figure*}

\begin{figure*}
  \includegraphics[scale=0.3]{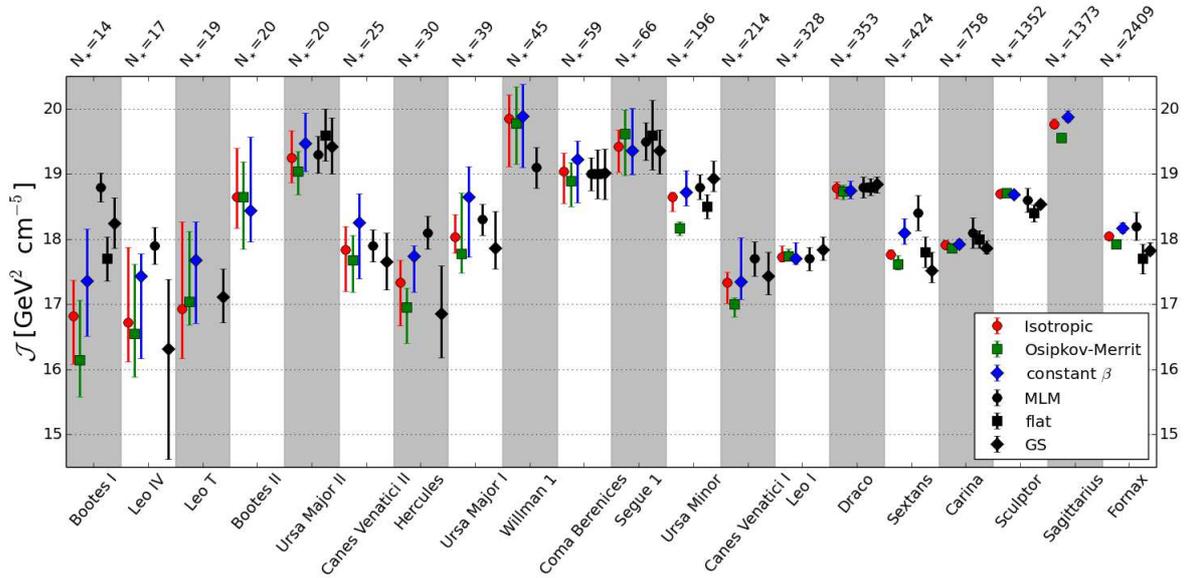}
 \caption{Results of the MLE scheme applied on the kinematic data from 20 dSphs using the 
  three models for the stellar velocity anisotropy considered in this study: isotropic 
  (red circles), constant $\beta$ (blue diamonds) and Osipkov-Merrit (green squares). 
  For all dSphs, the DM distribution and the stellar surface brightness in equation 
  \ref{eq4} corresponded to NFW and Plummer profiles. The black points refer to the 
  results reported in \citet[flat]{Ackermann:2011wa} (squares), 
  \citet[MLM]{Ackermann:2015zua} (cicles) and \citet[GS]{Geringer-Sameth:2014yza} 
  (diamonds) using Bayesian techniques (see text). The errorbars correspond to the 
  1-$\sigma$ uncertainties.}
\label{fig7}
 \end{figure*}
\indent To assess the statistical robustness of our method at different $N_\star$, we 
partition the full mock data set ($10^4$ stars) three times, into subsets containing 
either 10, 100, or 1000 stars. Note that this yields many (1000) subsets for testing 
$N_\star = 10$, while only 10 sets are available for $N_\star = 1000$. With these, first 
we assessed the bias and evaluated its uncertainty. In Fig.~\ref{fig5} we show the mean 
value of the recovered $\mathcal{J}_\text{MLE}$ for the eight models considered. In 
general the bias is smaller than 1 per cent. Notably, the bias is rather small even for 
$N_\star$ as low as 10, except for the case of an OM Cusped non-Plummer model. The 
somewhat larger bias for anisotropic models is likely due to the fact that while equation 
(\ref{eq3}) is a good approximation to the isotropic case it does not correspond to the 
true likelihood implemented to generate the simulated data. Next, we tested the coverage 
of the confidence intervals for each of the eight models and three sample sizes, and 
summarize the results in Fig.~\ref{fig6}. The green band represents the range of expected 
coverage estimates. Notably, the method seems to provide over-coverage (i.e. is 
conservative) for the case with only 10 stars. The coverage is overall satisfactory and 
the most critical cases of under-coverage occur when including the anisotropy parameter in 
the analysis of the set of ten samples with $N_\star = 1000$. Part of the undercoverage 
can be due to problems minimizing the large and degenerate parameter space that results 
from the inclusion of the anisotropy parameter (as for the methodological test the fitting 
for individual cases is not optimized) and the above mentioned aspects that the likelihood 
model we use is not entirely correct in the anisotropic case. In summary, despite the 
non-optimal modelling, the statistical properties are satisfactory and provide a strong 
indication of the reliability of the method and the statistical approach adopted.

The validation presented in this section was made assuming the true models used to create 
the mock data sets. Application of our method to real data incurs a new systematic 
uncertainty, because the true model is not known. The size of this uncertainty can be 
estimated by comparison with other methods, see, e.g. Fig.~\ref{fig7}. More directly, we 
can repeat our analysis while assuming a different model than the one used to generate 
the mock data. We do so, calculating the $\mathcal{J}$ factor for the small and large (100 
and 1000 stars) mock samples generated with the `Isotropic Cored Plummer-Like' model, 
but assuming the `Isotropic NFW Plummer' model, which we also use on real data 
(Sec.~\ref{sec4}). The results of this exercise are shown in Fig.~\ref{fig4}. We find the 
$\mathcal{J}$ factor to change by approximately 1 per cent, independent of the sample 
size. This uncertainty is not a characteristic of our frequentist method, and is rather a 
feature of the particular model space we are choosing. We expect this systematic to become 
smaller as more parameters of the DM profile are freed in the fit, and the model space 
becomes continuous.

\section{Results and Discussion}
\label{sec4}
In this Section we present the results from the application of our scheme using kinematic 
data from 20 dSphs. In order to properly compare to previous Bayesian-based methods, we 
use the same kinematic data as in the previous \textit{Fermi}-LAT 
analyses~\citep{Ackermann:2011wa,Ackermann:2015zua}. In both of these studies, the 
$\mathcal{J}$ factors and their uncertainties were obtained by sampling the posterior 
likelihood, using flat priors on the DM density profile parameters, in the former, and 
multi-level modelling (MLM) by \citet{2015MNRAS.451.2524M} in the latter. 

Our analysis assumes an NFW profile for the DM density, a Plummer profile for the surface 
brightness of the dSphs and one of the three velocity anisotropy profiles defined in 
Sec.~\ref{sec2}. Similar to the assumptions in the \textit{Fermi}-LAT analyses, we fix 
the angular integration angle to $\theta_\text{max} = 0.5^\circ$. For most of the 
brightest dSphs, this roughly corresponds to the half-light radius of the projected 
stellar distribution. It is within this approximate angular region that the integrated DM 
mass, and the J-factor, are most strongly constrained from the Jeans equations 
(\citealt{Wolf:2009tu}, \citealt{Walker:2011fs}). Furthermore, this is the region where 
most of the annihilation signal is expected to originate (\citealt{Strigari:2006rd}, 
\citealt{SanchezConde:2011ap}).

The MLE of $\mathcal{J}$ together with their 1-$\sigma$ uncertainties are presented in 
Table \ref{tab2}. In Fig.~\ref{fig7}, (where possible) we compare our results for all 
three velocity anisotropy models assumed (colored points) with previous results (black 
points) from \citet{Ackermann:2011wa,Ackermann:2015zua}, as well as 
from~\citet{Geringer-Sameth:2014yza}. The later results from 
\citet{Geringer-Sameth:2014yza} are similar to the flat prior case considered 
in~\citet{Ackermann:2011wa}, though the former marginalizes over a larger set of 
parameters that describe the DM profile. 

\setlength{\extrarowheight}{3pt}
\begin{table}
\caption{$\mathcal{J}$ factors derived from the kinematic data of 20 dSphs assuming an 
  isotropic (Iso), a constant anisotropy (Ca) and the Osipkov\---Merrit (OM) stellar 
  velocity anisotropy profiles, respectively. The uncertainties correspond to the 
  1$\sigma$ errors obtained from the likelihood curve constructed with MLE scheme we have 
  developed (see Sec.~\ref{sec2}). The values in the second column indicate the number of 
  stars in each data set.} 
\label{tab2}
\resizebox{\columnwidth}{!}{
\begin{tabular}{ l c c c c }
   \hline
            Dwarf &N$_\star$& $\mathcal{J}_\text{MLE}$(Iso) 	& 
$\mathcal{J}_\text{MLE}$(Ca)	 & $\mathcal{J}_\text{MLE}$(OM) \\
	    \hline \\
         Bootes I &    14   & $16.82_{-0.74}^{+0.55}$   	& $17.36_{-0.85}^{+0.79}$ 
 
	 &  $16.15_{-0.56}^{+0.92}$ \\
           Leo IV &    17   & $16.73_{-0.60}^{+1.15}$   	& $17.44_{-1.27}^{+0.34}$ 
 
	 &  $16.55_{-0.66}^{+1.07}$ \\
            Leo T &    19   & $16.93_{-0.76}^{+1.34}$   	& $17.68_{-0.97}^{+0.59}$ 
 
	 &  $17.04_{-0.35}^{+1.08}$ \\
    Ursa Major II &    20   & $19.26_{-0.38}^{+0.41}$   	& $19.47_{-0.43}^{+0.46}$ 
 
	 &  $19.05_{-0.36}^{+0.30}$ \\
        Bootes II &    20   & $18.64_{-0.47}^{+0.75}$   	& $18.44_{-0.48}^{+1.13}$ 
 
	 &  $18.65_{-0.80}^{+0.54}$ \\
Canes Venatici II &    25   & $17.83_{-0.63}^{+0.36}$   	& $18.25_{-0.86}^{+0.44}$ 
 
	 &  $17.68_{-0.49}^{+0.38}$ \\
         Hercules &    30   & $17.33_{-0.66}^{+0.35}$   	& $17.74_{-0.55}^{+0.16}$ 
 
	 &  $16.96_{-0.55}^{+0.29}$ \\
     Ursa Major I &    39   & $18.03_{-0.24}^{+0.34}$   	& $18.65_{-0.92}^{+0.46}$ 
 
	 &  $17.78_{-0.30}^{+0.93}$ \\
        Willman 1 &    45   & $19.86_{-0.74}^{+0.36}$   	& $19.89_{-0.79}^{+0.48}$ 
 
	 &  $19.77_{-0.62}^{+0.57}$ \\
   Coma Berenices &    59   & $19.05_{-0.49}^{+0.28}$   	& $19.22_{-0.66}^{+0.28}$ 
 
	 &  $18.90_{-0.40}^{+0.28}$ \\
          Segue 1 &    66   & $19.42_{-0.39}^{+0.26}$   	& $19.36_{-0.36}^{+0.66}$ 
 
	 &  $19.62_{-0.64}^{+0.37}$ \\
       Ursa Minor &   196   & $18.64_{-0.22}^{+0.08}$   	& $18.73_{-0.21}^{+0.33}$ 
 
	 &  $18.17_{-0.11}^{+0.10}$ \\
 Canes Venatici I &   214   & $17.33_{-0.32}^{+0.16}$   	& $17.35_{-0.27}^{+0.68}$ 
 
	 &  $17.01_{-0.20}^{+0.10}$ \\
            Leo I &   328   & $17.73_{-0.08}^{+0.17}$   	& $17.70_{-0.08}^{+0.25}$ 
 
	 &  $17.74_{-0.07}^{+0.11}$ \\
            Draco &   353   & $18.79_{-0.16}^{+0.09}$   	& $18.75_{-0.13}^{+0.15}$ 
 
	 &  $18.73_{-0.12}^{+0.10}$ \\
          Sextans &   424   & $17.76_{-0.06}^{+0.08}$   	& $18.10_{-0.18}^{+0.21}$ 
 
	 &  $17.61_{-0.08}^{+0.14}$ \\
           Carina &   758   & $17.91_{-0.05}^{+0.08}$   	& $17.95_{-0.08}^{+0.04}$ 
	
 &  $17.86_{-0.05}^{+0.07}$ \\
         Sculptor &  1352   & $18.70_{-0.04}^{+0.03}$   	& $18.68_{-0.03}^{+0.04}$ 
	
 &  $18.71_{-0.06}^{+0.05}$ \\
      Sagittarius &  1373   & $19.77_{-0.07}^{+0.08}$   	& $19.88_{-0.08}^{+0.10}$ 
 
	 &  $19.56_{-0.07}^{+0.06}$ \\
           Fornax &  2409   & $18.04_{-0.04}^{+0.03}$   	& $18.17_{-0.08}^{+0.08}$ 
 
	 &  $17.93_{-0.03}^{+0.05}$ \\
  \hline
\end{tabular}
}
\end{table}

We note that the uncertainties on the Bayesian results (i.e. the error bars of the black 
points) are reflective of both the statistical uncertainties and the assumed priors. This 
becomes evident when noting that the errors are weakly dependent on the size of the sample 
of stars. By contrast, for our method the confidence intervals decrease with the sample 
size, as expected from statistical uncertainties. For most dSphs, there is broadly good 
agreement with the previous Bayesian-derived results. The biggest deviations occur for 
Bootes 1, in particular in comparison to the MLM analysis. We believe that this is mostly 
due to the assumed prior in the MLM analysis. \\
\indent The overall spread in the best-fitting values for the $\mathcal{J}$ factors 
obtained from different methods illustrates the systematic impact of our modelling. The 
$\mathcal{J}$ factor is relatively insensitive to our model assumptions, with a maximum 
difference of 7 per cent for Bootes I. The overall spread between the different 
estimates can be taken as a rough indication of overall systematic uncertainties. In this 
particular comparison, it can be seen that the spread is roughly comparable to the 
statistical uncertainties for dwarfs with less than about 100 stars, whereas the 
systematic uncertainties dominate for the brightest dwarfs. \\

We reiterate that our results have been obtained assuming a particular model for the DM 
density, stellar and velocity anistropy profiles. This resulted in the reduction of the 
dimensionality of the full parameter space of the problem, hence in the size of the 
parameters array fit by MINUIT. Relaxing some of these assumptions and fitting a broader 
parameter space would allow the investigation of possible deviations from the model used, 
at the expense of reintroducing the degeneracy between the various profiles parameters. 
The effect of this modification and of the systematic effects arising from different 
model assumptions will be studied in a future project. Finally, though the likelihood 
function that we utilize does not lend itself to a straightforward definition of 
goodness-of-fit, we have checked that for all dSphs we study the best-fitting velocity 
dispersion profiles provide an acceptable $\chi^2$ per degree of freedom in comparison to 
the binned velocity dispersion data.

\section{Conclusions}
In this paper, we show that J-factor estimates and proper confidence intervals can be 
derived from stellar data with a completely frequentist Jeans analysis, based on a 
profile likelihood methodology. We validate this procedure with mock data sets from the 
\textit{Gaia} Challenge, and present prior-free estimates of the J-factors for 20 dSphs, 
removing the systematic uncertainty introduced by the often arbitrary choice of priors. 
Note that we do not attempt to compare the precision of our estimates with those in 
literature as this is only meaningful under similar model assumptions. We simply conclude 
that our method provides the desired statistical properties. In principle, there is no 
limitation to the extension of our method to more general models, depending on the 
envisaged application. Furthermore, this method opens the way to a correct definition of 
the likelihood function, essential for a self-consistent treatment of DM searches with 
gamma-ray data, which are predominantly performed in a frequentist framework. In future 
work we will update such gamma-ray analyses, provide a detailed study of the 
aforementioned systematic uncertainties compared to prior-based derivation, and extend 
the framework to non-Gaussian assumptions on velocity sampling distributions. Additional 
improvement may be obtained by appealing the measured distribution of Milky Way 
foreground stars \citep{Ichikawa:2016nbi}, which can be particularly important for dSphs 
with small kinematic samples. 

\section*{Acknowledgements}
AC is thankful to H. Silverwood and to M. Meyer for useful discussions. JC-T acknowledges 
the support of the Laboratoire de Physique Corpusculaire de Clermont-Ferrand, where part 
of this work was done. JC thanks support from  the Knut and Alice Wallenberg foundation, 
Swedish Research Council and Swedish National Space Board. LES acknowledges support from 
NSF grant PHY-1522717. BA acknowledges the support from the Swedish Research Council (PI: 
J. Conrad). MASC is a Wenner-Gren Fellow and acknowledges the support of the Wenner-Gren 
Foundations to develop his research. 


\bibliographystyle{mnras}
\bibliography{biblio}

\bsp	
\label{lastpage}
\end{document}